# FedsLLM: Federated Split Learning for Large Language Models over Communication Networks

Kai Zhao
College of Information Science and
Electronic Engineering
Zhejiang University
Hangzhou, China
2020302021200@whu.edu.cn

Zhaohui Yang*
College of Information Science and
Electronic Engineering
Zhejiang University
Hangzhou, China
yang_zhaohui@zju.edu.cn

Chongwen Huang
College of Information Science and
Electronic Engineering
Zhejiang University
Hangzhou, China
chongwenhuang@zju.edu.cn

Xiaoming Chen
College of Information Science
and Electronic Engineering
Zhejiang University
Hangzhou, China
chen_xiaoming@zju.edu.cn

Zhaoyang Zhang
College of Information Science
and Electronic Engineering
Zhejiang University
Hangzhou, China
ning_ming@zju.edu.cn

*Abstract*—Addressing the challenges of deploying large language models in wireless communication networks, this paper combines low-rank adaptation technology (LoRA) with the splitfed learning framework to propose the federated split learning for large language models (FedsLLM) framework. The method introduced in this paper utilizes LoRA technology to reduce processing loads by dividing the network into client subnetworks and server subnetworks. It leverages a federated server to aggregate and update client models. As the training data are transmitted through a wireless network between clients and both main and federated servers, the training delay is determined by the learning accuracy and the allocation of communication bandwidth. This paper models the minimization of the training delay by integrating computation and communication optimization, simplifying the optimization problem into a convex problem to find the optimal solution. Additionally, it presents a lemma that describes the precise solutions to this problem. Simulation results demonstrate that the proposed optimization algorithm reduces delays by an average of 47.63% compared to unoptimized scenarios.

## I. INTRODUCTION

With the increasing prevalence of mobile devices, there has been an explosive growth in the demand for wireless communication, leading to continuously expanding network capacity and traffic. A fundamental feature of 6G is the deep integration of AI with wireless networks, enhancing the delivery of intelligent services and applications. Machine learning plays a crucial role in supporting wireless communication services, achieving significant success in applications such as channel estimation [1], data detection, beamforming, and hybrid precoding. Facing the future's complex demands, including higher capacity access and massive data handling, large models offer viable solutions. The evolution of communication technology from traditional signal processing through conventional machine learning to future big wireless models is ongoing [2]. Leveraging extensive parameter scales and training data volumes, deploying large models in wireless networks enables the construction of intelligent networks capable of multitasking (multimodal), multi-scenario, and integrated scheduling.

Deploying large models in wireless communication networks presents several challenges. Large models typically have extensive parameters, requiring significant resources for fine-tuning across different downstream tasks. To address this, researchers have developed parameter-efficient fine-tuning techniques, such as Low-Rank Adaptation (LoRA) [3]. However, data distributed across various clients is often imbalanced, leading to the proposal of a federated learning (FL) framework to collaboratively train data across each client [4] [5].

Deploying large language models becomes challenging in environments with limited communication and computational resources, such as mobile and IoT devices. Researchers have proposed split learning (SL) [6] [7], where models are strategically partitioned and deployed across client and server ends to minimize processing loads on resource-constrained devices. Splitfed learning (SFL) combines the benefits of FL and SL [8], offering better model privacy than FL, and faster training speeds compared to SL, while maintaining similar model accuracy and communication efficiency. However, existing approaches have not fully leveraged the advantages of FL and SL in distributed scenarios for large language models, nor have they adequately considered the integration of LoRA with SFL. Thus, we propose the federated split learning for large language models (FedsLLM) to address these issues.

Moreover, in FedsLLM, the computational and communication delays are influenced by the local model learning accuracy, communication bandwidth allocation, and

```
Algorithm 1 FedsLLM(Fed server)
Fed server executes:
if n=0 then
    Initialize $\Delta \mathbf{w}_c^{(n)}$
    Send $\Delta \mathbf{w}_c^{(n)}$ to all K clients for ClientUpdate($\Delta \mathbf{w}_{c,k}^{(n)}$)
else
    for $k \in \mathcal{K}$ in parallel do
        $\mathbf{h}_{c,k}^{(n)} \leftarrow$ ClientBackprop($d\mathbf{A}_k^{(n)}$)
    end for
    Client-side global model updates: $\Delta \mathbf{w}_c^{(n+1)} = \Delta \mathbf{w}_c^{(n)} + \frac{1}{K}\sum_{k=1}^{K}\mathbf{h}_{c,k}^{(n)}$
    Send $\Delta \mathbf{w}_c^{(n+1)}$ to all K clients for ClientUpdate($\Delta \mathbf{w}_c^{(n)}$)
end if
```

the split ratio. Previous studies have explored optimizations in similar context. Researchers [9] optimized subchannel power control, and layer selection to minimize latency per iteration in Parallel Split Learning. Another study [10] within the hybrid federated split learning (HFSL) framework introduced a multi-objective optimization algorithm driven by predictive Generative Adversarial Networks (GAN), balancing training time and energy consumption. Yang et al. [11] focused on energy-efficient transmission and computational resource allocation in wireless networks under FL. However, previous works have not addressed the issue of minimizing delays specifically within the federated split learning framework.

The primary contribution of this paper is to provide a framework for optimizing FedsLLM on wireless communication networks. We model the minimization of training delays in FedsLLM by integrating computation and communication optimizations. The optimization problem is simplified into a convex problem. Moreover, we present the lemma related to the precise solutions of this optimization.

## II. MODEL FRAMEWORK

To address the communication overhead caused by the large number of parameters in large language models, this paper employs a parameter-efficient fine-tuning technique known as LoRA. This technique maintains the original parameters of the large language model unchanged. Specifically, LoRA introduces a bottleneck module used solely for fine-tuning, which forms a residual connection with the original parameters $\boldsymbol{\omega}_0$. This bottleneck module consists of two matrices, *A* and *B*. Matrix *A* reduces the input dimension from $d$ to $r$, and matrix *B* restores the output dimension from $r$ back to $k$, as illustrated in Equation (1).

$$\boldsymbol{\omega}_0 + \Delta\boldsymbol{\omega} = \boldsymbol{\omega}_0 + BA, \\ s.t. B \in \mathcal{R}^{d \times r}, A \in \mathcal{R}^{r \times k}, r \ll \min(d,k). \quad (1)$$

FedsLLM leverages the primary advantages of FL and SL, enabling parallel processing across distributed clients while segmenting the network into client-side and server-side subnetworks during training, as illustrated in Figure.1. The fed server performs the FedAvg aggregation algorithm on local updates at the client-side, synchronizing the global model across clients in each training round. The specific workflow of FedsLLM is outlined in Algorithm 1. All clients perform forward propagation in parallel on their local models, including the noise layer, and send the scrambled data output to the main server. Subsequently, the main server processes this smashed data in parallel through forward and backward propagation for each client's server-side local model. After processing, the gradients of the smashed data are sent back to the corresponding clients for backward propagation. This process continues until the local models achieve a specified accuracy. Then, using the FedAvg algorithm, the main server updates its global model. Clients send their local model parameters to the fed server, which performs a weighted average update of the global model for all clients. The updated global client model is then broadcast back to each client via wireless transmission.

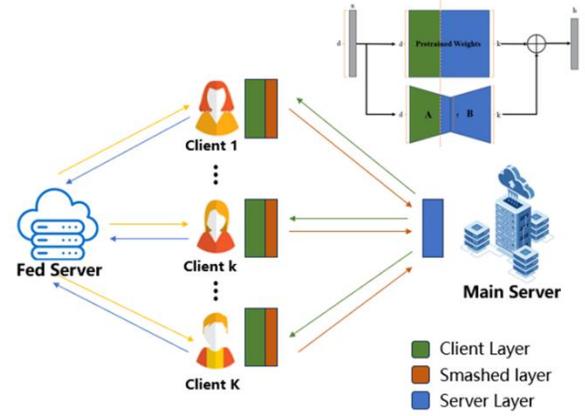

Fig. 1. FedsLLM framework

## III. MINIMIZING TRAINING DELAY

In this wireless network, composed of a fed server and K users, each user $k$ has a local dataset $\mathcal{D}_k$ with $D_k$ data samples, where each dataset $\mathcal{D}_k = \{\boldsymbol{x}_{kl}, y_{kl}\}_{l=1}^{D_k}$, with $\boldsymbol{x}_{kl} \in \mathbb{R}^d$ being the input vector for user $k$, and $y_{kl}$ being the corresponding output. The trainable parameters of the entire model $\Delta\boldsymbol{\omega}$ are divided into client model parameters $\boldsymbol{\omega}_c$ and server-side model parameters $\boldsymbol{\omega}_s$, which are deployed on the client and main server, respectively. The vector $\boldsymbol{\omega}_c$ represents the global model parameters trained by the client's dataset, while $\boldsymbol{\omega}_s$ represents the global model parameters trained by the main server's dataset. We introduce a loss function $f(\boldsymbol{\omega}_s, \boldsymbol{\omega}_c, \boldsymbol{x}_{kl}, y_{kl})$ to describe the performance of the model with input vector $\boldsymbol{x}_{kl}$ and output $y_{kl}$. The loss function varies depending on the learning task. The total loss function for the user $k$ is :

$$F_k(\boldsymbol{\omega}_0, \Delta\boldsymbol{\omega}) = \frac{1}{D_k}\sum_{l=1}^{D_k} f(\boldsymbol{\omega}_0, \boldsymbol{\omega}_s, \boldsymbol{\omega}_c, \boldsymbol{x}_{kl}, y_{kl}). \quad (2)$$

The FedsLLM training problem can be formulated as follows:

$$\min_{\Delta\boldsymbol{\omega}} F(\boldsymbol{\omega}_0, \Delta\boldsymbol{\omega}) = \frac{1}{D}\sum_{k=1}^{K}\sum_{l=1}^{D_k} f(\boldsymbol{\omega}_0, \boldsymbol{\omega}_s, \boldsymbol{\omega}_c, \boldsymbol{x}_{kl}, y_{kl}). \quad (3)$$

where $D = \sum_{k=1}^{K} D_k$ is the total data samples of all users.

In this paper, we define the vector $\Delta\boldsymbol{\omega}^{(n)}$ as the combination of the global model parameters of the client and server for the

**Algorithm 2** FedsLLM(Main Server and Client)

---
**Main Server executes:**
**if** n=0 **then**
    Initialize $\Delta \mathbf{w}_s^{(n)}$
    **for** $k \in \mathcal{K}$ in parallel **do**
        **repeat**
            $(\mathbf{A}_k^{(n)}, \mathbf{y}_k) \leftarrow ClientUpdate(\Delta \mathbf{w}_c^{(n)}, \mathbf{h}_{c,k}^{(n)})$
            Forward propagation with $\mathbf{A}_k^{(n)}$ on $\Delta \mathbf{w}_s^{(n)}$ and $\mathbf{h}_{s,k}^{(n)}$ to compute $\hat{\mathbf{y}}_k$
            Loss calculation with $\mathbf{y}_k$ and $\hat{\mathbf{y}}_k$
            Back-propagation calculate $\nabla G_k(\Delta \mathbf{w}_s^{(n)}; \mathbf{A}_k^{(n)}; \mathbf{h}_{s,k}^{(n)})$
            Update $\mathbf{h}_{s,k}^{(n),(i+1)} = \mathbf{h}_{s,k}^{(n),(i)} - \delta_s \nabla G_k\left(\mathbf{w}_s^{(n)}, \mathbf{h}_{s,k}^{(n),(i)}\right)$
            Send $d\mathbf{A}_k^{(n)} := \nabla G_k(\mathbf{A}_k^{(n)}; \Delta \mathbf{w}_s^{(n)}; \mathbf{h}_{s,k}^{(n)})$ to client k for ClientBackprop($d\mathbf{A}_k^{(n)}$)
        **until** client k solve problem(4) with a given learning accuracy $\eta$ and the solution is $\mathbf{h}_{s,k}^{(n)}$
    **end for**
    Main Server model update: $\Delta \mathbf{w}_s^{(n+1)} = \Delta \mathbf{w}_s^{(n)} + \frac{1}{K} \sum_{k=1}^{K} \mathbf{h}_{s,k}^{(n)}$
**end if**

**Client k executes:**
**ClientUpdate**($\Delta \mathbf{w}_c^{(n)}$):
Model updates $\mathbf{w}_c^{(n)} \leftarrow$ FedServer()
Initialize $\mathbf{h}_{c,k}^{(n)}$
**repeat**
    Forward propagation with data $\mathbf{x}_k$ up to the final layer
    $\mathbf{y}_k$ is the true labels of $\mathbf{x}_k$
    $\mathbf{A}_k^{(n)}$ is the outputs of the final layer
    Send $\mathbf{y}_k$ and $\mathbf{A}_k^{(n)}$ to the main server
    Wait for the completion of ClientBackprop($d\mathbf{A}_k^{(n)}$)
**until** client k solve problem(4) with a given learning accuracy $\eta$ and the solution is $\mathbf{h}_{c,k}^{(n)}$

**ClientBackprop**($d\mathbf{A}_k^{(n)}$):
**repeat**
    Back-propagation,calculate gradients $\nabla G_k(\Delta \mathbf{w}_c^{(n)}; \mathbf{h}_{c,k}^{(n)})$ with $d\mathbf{A}_k^{(n)}$
    Update $\mathbf{h}_{c,k}^{(n),(i+1)} = \mathbf{h}_{c,k}^{(n),(i)} - \delta_c \nabla G_k\left(\mathbf{w}_c^{(n)}, \mathbf{h}_{c,k}^{(n),(i)}\right)$
**until** client k solve problem(4) with a given learning accuracy $\eta$ and the solution is $\mathbf{h}_{c,k}^{(n)}$
Send $\mathbf{h}_{c,k}^{(n)}$ to the Fed server

---

nth global iteration, namely $\boldsymbol{\omega}_c^{(n)}$ and $\boldsymbol{\omega}_s^{(n)}$. The local problem computed by user $k$ and the main server is :

$$\min_{\mathbf{h}_k \in \mathbb{R}^d} G_k(\Delta \boldsymbol{\omega}_x^{(n)}, \mathbf{h}_{x,k}) \triangleq F_k(\Delta \boldsymbol{\omega}_x^{(n)} + \mathbf{h}_{x,k})$$
$$-(\nabla F_k(\Delta \boldsymbol{\omega}_x^{(n)}) - \xi \nabla F(\Delta \boldsymbol{\omega}_x^{(n)}))^T \mathbf{h}_{x,k} \quad (4)$$

where x belongs to {c, s} ,and $\xi$ is a constant. The vector $\mathbf{h}_k$ represents the solution for local model parameter updates by user $k$ and the main server during each local iteration, consisting of $\mathbf{h}_{c,k}$ for the client and $\mathbf{h}_{s,k}$ for the main server. Specifically, $\mathbf{w}_{c,k}^{(n)} + \mathbf{h}_{c,k}^{(n)}$ describes the parameters of the local client model of user $k$ at the $n_{th}$ global iteration, and $\mathbf{w}_{s,k}^{(n)} + \mathbf{h}_{s,k}^{(n)}$ corresponds to the parameters of the local model at the main server for user $k$. Since obtaining an exact solution for Problem (4) is typically challenging, an approximate solution with a defined accuracy $\eta$ is used. The solution within the accuracy $\eta$ at the $n_{th}$ iteration is :

$$G_k(\Delta \boldsymbol{\omega}^{(n)}, \mathbf{h}_k^{(n)}) - G_k(\Delta \boldsymbol{\omega}^{(n)}, \mathbf{h}_k^{(n)*})$$
$$\leq \eta \left( G_k(\Delta \boldsymbol{\omega}^{(n)}, \mathbf{0}) - G_k(\Delta \boldsymbol{\omega}^{(n)}, \mathbf{h}_k^{(n)*}) \right), \quad (5)$$

where $\mathbf{h}_k^{(n)*}$ is the actual optimal solution of problem (4).

In the $n_{th}$ iteration, the solution for problem (3.2) within the accuracy $\epsilon_0$ can be :

$$F(\Delta \boldsymbol{\omega}^{(n)}) - F(\Delta \boldsymbol{\omega}^*) \leq \epsilon_0 \left( F(\Delta \boldsymbol{\omega}^{(0)}) - F(\Delta \boldsymbol{\omega}^*) \right), \quad (6)$$

where $\Delta \boldsymbol{\omega}^*$ is the exact solution for problem (3).

To analyze the convergence of Algorithm 1, we assumes that $F_k(\Delta \boldsymbol{\omega})$ is $L$-Lipschitz continuous and $\gamma$-strongly convex, which is expressed as:

$$\gamma \mathbf{I} \leq \nabla^2 F_k(\Delta \boldsymbol{\omega}) \leq L\mathbf{I}, \forall k \in \mathcal{K}. \quad (7)$$

Under assumption (7), the paper references lemma 1[11] concerning the convergence rate of Algorithm 1.

**Lemma 1** For Algorithm 1, if $0 < \xi \leq \frac{\gamma}{L}$, then the solution $\Delta \boldsymbol{\omega}^{(n)}$ within the accuracy $\epsilon_0$ can be achieved when the number of global iterations

$$n \geq \frac{a}{1-\eta} \triangleq I_0, a = \frac{2L^2}{\gamma^2 \xi} \ln \frac{1}{\epsilon_0}, \quad (8)$$

It is observed that the required number of global iterations $n$ decreases as the accuracy $\epsilon_0$ in problem (3) increases, and increases as the accuracy $\eta$ in problem (4) increases.

The training process of FedsLLM includes local computations at client model, local computations at the main server model, parameter transmission between clients and the main server, parameter uploads from clients to the fed server, and the aggregation and broadcasting of the global model by the fed server. The delay modeling presented here assumes that no privacy protection measures such as noise layers or differential privacy are implemented in FedsLLM.

*A. Local computation*

we employs the gradient descent method to address Problem (4). The parameter update for the $i+1$ local iteration is calculated as:

$$\mathbf{h}_{x,k}^{(n),(i+1)} = \mathbf{h}_{x,k}^{(n),(i)} - \delta \nabla G_k(\Delta \boldsymbol{\omega}_x^{(n)}, \mathbf{h}_{x,k}^{(n),(i)}), x \in \{c, s\} \quad (9)$$

where $\delta$ represents the step size, $\mathbf{h}_k^{(n),(i)}$ denotes the solution for the local model parameter updates for user $k$ and the main server at the $i$-th local iteration,and The $\nabla G_k(\Delta \boldsymbol{\omega}_x^{(n)}, \mathbf{h}_{x,k}^{(n),(i)})$ is the gradient of the function $G_k(\Delta \boldsymbol{\omega}_x^{(n)}, \mathbf{h}_{x,k})$ at point $\mathbf{h}_{x,k} = \mathbf{h}_{x,k}^{(n),(i)}$. We set $\mathbf{h}_{x,k}^{(n),(0)} = 0$.

**Lemma 2** if $\delta < \frac{2}{L}$ and each client runs gradient descent for a number of local iterations $i \geq v \log_2(1/\eta)$, where $v = \frac{2}{(2-L\delta)\delta\gamma}$, then a solution within the accuracy $\eta$ for Problem (4) can be achieved.

The proof of Lemma 2 can be found in the literature [11].The total model is split into two parts, deployed separately on the client and the main server. Let $A$ represent the proportion of client model parameters to the total model parameters, then $(1 - A)$ represents the proportion of computational load on the main server model.Assuming the CPU frequency for user $k$ is

$f_k$ and for the main server is $f_s$, the computational time required for the client model and the main server model to achieve the specified accuracy $\eta$ is:

$$\tau = \tau_k + \tau_s = \frac{Av|\boldsymbol{\omega}_0 + \Delta\boldsymbol{\omega}|CD_k \log_2\left(\frac{1}{\eta}\right)}{f_k}$$
$$+ \frac{(1-A)v|\boldsymbol{\omega}_0 + \Delta\boldsymbol{\omega}|CD_k \log_2\left(\frac{1}{\eta}\right)}{f_s}$$
$$= E_k \log_2\left(\frac{1}{\eta}\right)\left(\frac{A}{f_k} + \frac{(1-A)}{f_s}\right), \forall k \in \mathcal{K}, \quad (10)$$

where $C$ represents the number of CPU cycles required to process a single data sample per parameter in the total model, $|\boldsymbol{\omega}_0 + \Delta\boldsymbol{\omega}|$ denotes the total number of parameters in the total model. According to Lemma 2, $v \log_2\left(\frac{1}{\eta}\right)$ is the minimum number of local iterations required for each user. In Equation (10), $E_k = v|\boldsymbol{\omega}_0 + \Delta\boldsymbol{\omega}|CD_k$.

## B. WIRELESS TRANSMISSION

After local computations, each user uploads their local model parameters or intermediate layer outputs to the fed server or main server via wireless transmission. The upload rate for user $k$ to the fed server or main server can be expressed as:

$$r_{x,k} = b_{x,k} \log_2\left(1 + \frac{g_{x,k}p_{x,k}}{N_x b_{x,k}}\right), x \in \{c, s\}, \forall k \in \mathcal{K}, \quad (11)$$

where $b_{x,k}$ represents the bandwidth allocated to user $k$, $p_{x,k}$ is the transmission power of user $k$ for uploading to the fed or main server, $N_x$ is the power spectral density of Gaussian white noise, and $g_{x,k}$ is the channel gain between user $k$ and the fed or main server. Due to the limited total bandwidth:

$$\sum_{k=1}^{K} b_{x,k} \leq B_x, x \in \{c, s\}, \quad (12)$$

where $B_x$ is the total bandwidth for uploads from clients to the fed or main server.

During this process, each user uploads the vector $\boldsymbol{h}_{x,k}^{(n),(i)}$ to the fed server. Since the dimension of vector $\boldsymbol{h}_{x,k}^{(n),(i)}$ is fixed for all clients, the data volume required per user per round is a constant denoted by $s_c$. To ensure that the data volume $s_c$ is transmitted within the upload time $t_{c,k}$ to the fed server, the following condition must be met:

$$t_{c,k} r_{c,k} \geq s_c, \quad (13)$$

Since the output data dimension of the last layer of the client model is fixed for all clients, the data volume that each user needs to upload to the main server per round is a constant, denoted as $s$. To ensure that the data volume $s$ is transmitted within the upload time $t_{s,k}$ to the main server, the following condition must be met:

$$t_{s,k} r_{s,k} \geq s. \quad (14)$$

## C. Aggregation and Broadcast

In this stage, the fed server collects local model parameters from various clients, updates the global model, and then broadcasts it back to all clients. Due to its high transmission power and substantial bandwidth, the fed server can broadcast the model quickly, making the broadcasting duration effectively negligible. The fed server, being a dedicated high-performance server, is capable of rapidly aggregating models, which allows us to overlook the aggregation time in practical terms.

Upon receiving the output data from clients, the main server proceeds with forward propagation, calculates the loss function, and initiates backpropagation. It transmits the gradient data from the final layer of the client models to the respective clients via a wireless network for further backpropagation. Given the main server's robust transmission capabilities and the significant bandwidth available, the transmission of gradients is accomplished swiftly, thus the duration is considered negligible.

## D. Problem Formulation

Hence, each training round's latency encompasses the local compute delays on both client and main server models, the upload time to the main server, and the upload time to the fed server. As defined by Equation (10), (13) and (14), the total latency for user $k$ across the client and main server models is calculated as:

$$T_k = I_0 \left(\tau + t_{c,k} + v \log_2\left(\frac{1}{\eta}\right) t_{s,k}\right)$$
$$= \frac{a}{1-\eta}\left(\begin{array}{c} E_k \log_2\left(\frac{1}{\eta}\right)\left(\frac{A}{f_k} + \frac{(1-A)}{f_s}\right) \\ + t_{c,k} + v \log_2\left(\frac{1}{\eta}\right) t_{s,k} \end{array}\right). \quad (15)$$

where $v \log_2\left(\frac{1}{\eta}\right)$ represents the minimum number of local iterations required per global cycle.

Define $T = \max_{k \in \mathcal{K}} T_k$ as the overall latency for the FedsLLM network's training algorithm.

We now pose the delay minimization problem:

$$\min_{T, t_c, t_s, b_c, b_s, f, p_c, p_s, f_s, \eta, A} T \quad (16)$$

s.t. $\frac{a}{1-\eta}\left(\begin{array}{c} E_k \log_2\left(\frac{1}{\eta}\right)\left(\frac{A}{f_k} + \frac{(1-A)}{f_s}\right) \\ + t_{c,k} + v \log_2\left(\frac{1}{\eta}\right) t_{s,k} \end{array}\right) \leq T, \forall k \in \mathcal{K}, \quad (16a)$

$$t_{s,k} b_{s,k} \log_2\left(1 + \frac{g_{s,k}p_{s,k}}{N_s b_{s,k}}\right) \geq s, \forall k \in \mathcal{K}, \quad (16b)$$

$$t_{c,k} b_{c,k} \log_2\left(1 + \frac{g_{c,k}p_{c,k}}{N_c b_{c,k}}\right) \geq s_c, \forall k \in \mathcal{K}, \quad (16c)$$

$$\sum_{k=1}^{K} b_{s,k} \leq B_s, \quad (16d)$$

$$\sum_{k=1}^{K} b_{c,k} \leq B_c, \quad (16e)$$

$$0 \leq f_k \leq f_k^{\max}, 0 \leq f_s \leq f_s^{\max},$$
$$0 \leq p_{s,k} \leq p_{s,k}^{\max}, 0 \leq p_{c,k} \leq p_{c,k}^{\max}, \forall k \in \mathcal{K}, \quad (16f)$$
$$0 \leq \eta \leq 1, A_{min} < A < A_{max} \quad (16g)$$
$$t_{s,k} \geq 0, b_{s,k} \geq 0, t_{c,k} \geq 0, b_{c,k} \geq 0, \forall k \in \mathcal{K}. \quad (16h)$$

where $\boldsymbol{t}_c = [t_{c,1}, \cdots, t_{c,K}]^T$ and $\boldsymbol{t}_s = [t_{s,1}, \cdots, t_{s,K}]^T$ represent the vectors of transmission times to the main server and fed server respectively for all clients, $\boldsymbol{b}_c = [b_{c,1}, \cdots, b_{c,K}]^T$ and $\boldsymbol{b}_s = [b_{s,1}, \cdots, b_{s,K}]^T$ are the bandwidth vectors allocated to clients for transmissions to the main and fed servers respectively, $\boldsymbol{f} = [f_1, \cdots, f_K]^T$ denotes the vector of CPU frequencies for all clients, $\boldsymbol{p}_s = [p_{s,1}, \cdots, p_{s,K}]^T$ and $\boldsymbol{p}_c = [p_{c,1}, \cdots, p_{c,K}]^T$ are the transmission power vectors for clients to the main and fed servers respectively. $f_k^{\max}$ is the maximum local computational capability for client $k$, $f_s^{\max}$ is the maximum local computational capability for the main server. $p_{s,k}^{\max}$ and $p_{c,k}^{\max}$ are the maximum transmission powers for client $k$ uploading to the main server and fed server respectively. $A_{min}$ and $A_{max}$ represent the minimum and maximum proportions of client model parameters relative to the complete model.

Constraints (16a) indicates that the total computation and transmission time of all clients with the main server must not exceed the overall delay of the FedsLLM algorithm. The data transmission time constraints are given by constraints (16b)-(16c), bandwidth allocation constraints for each client by constraints (16d)-(16e), and the maximum local computational capabilities and maximum transmission powers of all clients, as well as the maximum local computational capability of the main server, are provided in constraints (16f). Constraints (16g) sets the constraints for the range of the accuracy and partition ratio, and constraints (16h) imposes non-negativity constraints on the related parameters.

### E. Optimal Resource Allocation

Let $(\boldsymbol{t}_c^*, \boldsymbol{t}_s^*, \boldsymbol{b}_c^*, \boldsymbol{b}_s^*, \boldsymbol{f}^*, \boldsymbol{p}_c^*, \boldsymbol{p}_s^*, f_s^*, \eta^*, A^*)$ be the optimal solution to problem (16). From constraint (16a), it is observed that the left-hand side of the inequality concerning $f_k$ and $f_s$ is an increasing function. Therefore, to meet the minimum latency constraints, the local computing capacities of clients and the main server should be set to their maximum values, i.e., $f_k^* = f_k^{max}, f_s^* = f_s^{max}, \forall k \in \mathcal{K}$. Through constraints (16b), (16c), and (16f), it can be seen that latency is a decreasing function concerning $p_{s,k}$ and $p_{c,k}$. Thus, to minimize latency, the transmission power of client $k$ to the main server and fed server should be set to the maximum, i.e., $p_{c,k}^* = p_{c,k}^{\max}, p_{s,k}^* = p_{s,k}^{\max}, \forall k \in \mathcal{K}$.

Given that the main server typically has higher computational resources than the client devices, which may be resource-constrained IoT or mobile devices, the inequality $f_s^{\max} > f_k^{\max}$ holds true for all clients, which leads to $\left(\frac{1}{f_k^{\max}} - \frac{1}{f_s^{\max}}\right) > 0$. Thus, the right side of the inequality (16a) is a decreasing function with respect to $A$ and to maximize the right side of the inequality (16a), $A^* = A_{min}$ is chosen.

Incorporating the above optimization results into problem (16), the simplified latency minimization problem is obtained:

$$\min_{T, \eta, \boldsymbol{t}_c, \boldsymbol{t}_s, \boldsymbol{b}_c, \boldsymbol{b}_s} T \quad (17)$$

s.t. $\frac{a}{1-\eta}\left(E_k \log_2\left(\frac{1}{\eta}\right)\left(\frac{A_{min}}{f_k^{\max}} + \frac{(1-A_{min})}{f_s^{\max}}\right)\right)$
$+t_{c,k} - v \log_2(\eta) t_{s,k}$
$\leq T, \forall k \in \mathcal{K}, \quad (17a)$

$$\frac{s}{t_{s,k}} \leq b_{s,k} \log_2\left(1 + \frac{g_{s,k} p_{s,k}^{\max}}{N_s b_{s,k}}\right), \forall k \in \mathcal{K}, \quad (17b)$$

$$\frac{s_c}{t_{c,k}} \leq b_{c,k} \log_2\left(1 + \frac{g_{c,k} p_{c,k}^{\max}}{N_c b_{c,k}}\right), \forall k \in \mathcal{K}, \quad (17c)$$

$$\sum_{k=1}^{K} b_{s,k} \leq B_s, \quad (17d)$$

$$\sum_{k=1}^{K} b_{c,k} \leq B_c, \quad (17e)$$

$$0 \leq \eta \leq 1, \quad (17f)$$
$$t_{s,k} \geq 0, b_{s,k} \geq 0, t_{c,k} \geq 0, b_{c,k} \geq 0, \forall k \in \mathcal{K}. \quad (17g)$$

Consider the function $y = x \ln\left(1 + \frac{1}{x}\right)$ defined for $x > 0$. The derivatives are calculated as:

$$y' = \ln\left(1 + \frac{1}{x}\right) - \frac{1}{x+1}, y'' = -\frac{1}{x(x+1)^2} < 0. \quad (18)$$

This shows that $y'$ decreases with increasing $x$ and $\lim_{x \to +\infty} y' = 0$, confirming that $y'$ is positive for all $x > 0$. Therefore, $y$ is an increasing function over its domain. Consequently, the right sides of the inequalities (17b) and (17c) increase with $b_{s,k}$ and $b_{c,k}$, respectively. Given the constraint on maximum bandwidth, it is preferable to minimize $b_{s,k}$ and $b_{c,k}$, maximizing $t_{s,k}$ and $t_{c,k}$ as implied by inequality (17a) to achieve minimal delay. Based on the constraints (17a), (17b), and (17c), Lemma 3 is derived.

**Lemma 3** The exact solutions for the transmission times $t_{s,k}$ and $t_{c,k}$ must satisfy the following condition for all $k \in \mathcal{K}$:

$$t_{c,k}^* + v \log_2\left(\frac{1}{\eta}\right) t_{s,k}^* = \frac{(1-\eta)T}{a}$$
$$+ E_k \log_2 \eta \left(\frac{1}{f_s^{\max}} + \left(\frac{1}{f_k^{\max}} - \frac{1}{f_s^{\max}}\right) A_{min}\right), \forall k \in \mathcal{K}, \quad (19)$$

The precise solutions for $b_{s,k}$ and $b_{c,k}$ should satisfy:

$$\frac{s}{t_{s,k}^*} = b_{s,k}^* \log_2\left(1 + \frac{g_{s,k} p_{s,k}^{\max}}{N_s b_{s,k}^*}\right), \forall k \in \mathcal{K}, \quad (20)$$

$$\frac{s_c}{t_{c,k}^*} = b_{c,k}^* \log_2\left(1 + \frac{g_{c,k} p_{c,k}^{\max}}{N_c b_{c,k}^*}\right), \forall k \in \mathcal{K}, \quad (21)$$

Given a specific value of $\eta$, problem (17) features linear inequality constraints in constraints (17a), (17d), and (17e), which collectively form a convex set. The inequalities in constraints (17b) and (17c) involve convex functions on their left sides and concave functions on their right sides, establishing a convex feasible set overall. Hence, problem (17) under a specified $\eta$ qualifies as a convex optimization problem.

To solve the latency minimization challenge, the approach involves systematically varying $\eta$ from 0 to 1 in predetermined increments. For each value of $\eta$, the corresponding convex optimization problem (17) is solved. The optimal solution $T^*$ is determined by selecting the minimum latency value from all these solutions, with the associated $\eta$ designated as $\eta^*$. This method ensures that the latency is minimized effectively within the constraints set by the problem.

## IV. SIMULATION RESULTS

This study conducted simulation experiments involving 50 users evenly distributed in a 500m x 500m square area, centered around a base station. The users uploaded their local parameters using Frequency Division Multiple Access (FDMA), adopting a path loss model of $128.1 + 37.6\log_{10}(d)$, with $d$ representing the distance in kilometers and shadow fading having a standard deviation of 8dB. The noise power spectral density was set at $N_0 = -174$ dBm/Hz. A real dataset from an open blog feedback platform [12] was utilized, comprising 60,021 blog post samples, each with 281 dimensions. Each data sample required $[1,3] \times 10^4$ cycles, distributed uniformly across the parameter $C_k$. Local computations were characterized by an effective switching capacitance $\kappa = 10^{-28}$. The settings for Algorithm 1 included $\xi = 1/10, \delta = 1/10$, and $\epsilon_0 = 10^{-3}$. Default settings were used for the maximum transmission power and computational capacity of all users, set at 10dBm and 2GHz respectively. Each data sample required $s_c = 28.1$ kbits for transmission, while each round of local iteration demanded $s = 281$kbits uploaded to the main server, with a total bandwidth 20MHz for uploads communication. Each user had an equal probability of selecting samples from the dataset. Simulations were performed in the MATLAB environment, systematically iterating $\eta$ from 0 to 1 in increments of 0.01, employing the interior-point method of the fmincon function to tackle the convex optimization problem posed.

In this study, we conducted comparative experiments among the proposed FedsLLM resource allocation strategy and three alternative FedsLLM strategies: one using equal bandwidth with a focus solely on optimizing η (labelled as 'EB'), another fixing the local iteration accuracy at $\eta = 0.1$ with a focus on optimizing bandwidth allocation (labelled as 'FE'), and the third fixing η at 0.1 and using equal bandwidth (labelled as 'BA'). The outcomes are depicted in Fig. 2, illustrating the variation in training latency as a function of each user's maximum average transmission power. The results clearly show that the proposed approach outperforms the others, achieving a significant reduction in latency—approximately 47.63% lower on average than the BA strategy. This superior performance stems from the method's comprehensive optimization of both bandwidth allocation and the precision parameter $\eta$, unlike the EB strategy where bandwidth is fixed, the FE strategy which does not optimize η, and the BA strategy where neither parameter is optimized.

## V. CONCLUSION

This study focuses on the deployment of large language models in environments where resources are limited, proposing the FedsLLM framework. It addresses the challenge of minimizing latency in the FedsLLM framework within wireless communication networks by reformulating this challenge as a convex optimization problem, thereby facilitating the derivation of an optimal solution. The simulation outcomes clearly indicate that the algorithm developed under this framework significantly enhances performance compared to scenarios where no optimization is applied.

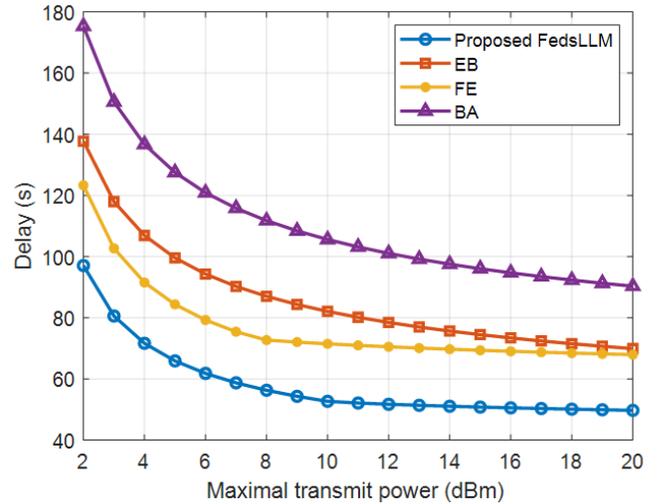

Fig. 2. Client's Minimum Training Latency Across Varying Levels of Maximum Transmission Power